\documentstyle[12pt,epsf]{article}

\newcommand{\pl}[3]{Phys.\ Lett.\ {\bf #1B}, #2 (19#3)}
\newcommand{\np}[3]{Nucl.\ Phys.\ {\bf B#1}, #2 (19#3)}
\newcommand{\prd}[3]{Phys.\ Rev.\ {\bf D#1}, #2 (19#3)}
\newcommand{\prl}[3]{Phys.\ Rev.\ Lett.\ {\bf #1}, #2 (19#3)}

\newcommand{\beq}{\begin{equation}}
\newcommand{\eeq}{\end{equation}}
\newcommand{\bea}{\begin{eqnarray}}
\newcommand{\eea}{\end{eqnarray}}
\renewcommand{\a}{\alpha}
\newcommand{\s}{\sigma}
\newcommand{\hsig}{\hat \sigma}
\newcommand{\hs}{\hat s}
\newcommand{\kt}{k_\perp}
\newcommand{\ra}{\rightarrow}
\newcommand{\nn}{\nonumber}

\begin{document}
 
\begin{titlepage}
 
\hspace*{\fill}\parbox[t]{3.8cm}{DESY 96-220\\ANL-HEP-PR-96-46\\
Edinburgh 96/26 \\ October 1996 
\\ hep-ph 96?}
 
\vspace*{1cm}
 
\begin{center}
\large\bf
Azimuthal Dependence of Forward-Jet Production in DIS in the High-Energy Limit
\end{center}
 
\vspace*{0.5cm}
 
\begin{center}
Jochen Bartels$^1$, Vittorio Del Duca$^2$ and Mark W\"usthoff$^3$
\end{center}
\begin{center} 
$^1$II.Institut f\"ur Theoretische Physik, Universit\"at Hamburg, \\
D-22761 Hamburg, Germany\\
$^2$Dept. of Physics and Astronomy, University of Edinburgh,\\
Edinburgh EH9 3JZ, Scotland, U.K.\\
$^3$ High Energy Physics Division,
Argonne National Laboratory, USA\footnote{Supported by the
U.S. Department of Energy, Contract W-31-109-ENG-38}.
\end{center}
 
\vspace*{0.5cm}
 
\begin{center}
\bf Abstract:
\end{center}
\noindent
As a signal for the BFKL Pomeron in small-x deep inelastic $ep$ scattering, we
calculate the azimuthal dependence of the inclusive cross section of forward
jets relative to the outgoing electron. For not very large differences in 
rapidity
between the current jet and the forward jet the cross section peaks at
$\pi/2$. For increasing rapidity BFKL dynamics predicts a decorrelation
in the azimuthal dependence between the electron and the forward jet.     
\end{titlepage}
 
\baselineskip=0.8cm
 
\section{Introduction}
Forward jet production in Deep Inelastic Scattering is of great value for
testing QCD at large energies where the
probability of having  multi-gluon emission is strongly increased and 
the need arises of resumming diagrams to all orders in $\alpha_s$.
The resummation can only be performed to leading log accuracy
and is technically done by solving the BFKL-equation \cite{BFKL}.  
The virtuality $Q^2$ and the jet transverse momentum $k_t^2$, both of them 
being larger than $5GeV^2$, provide enough hardness in the process to
justify the use of perturbative QCD excluding soft
contributions.
At fixed $Q^2$ the maximum parton subenergy $\hs$ is achieved by taking the 
lowest possible $x_{bj}$-values and the largest parton or jet energy $xp$ 
with $p$ being the proton momentum.

In hadron-hadron collisions, kinematic configurations of this kind may be
realized by selecting two-jet events at large rapidity intervals 
$\eta\simeq\ln(\hs/Q^2)$, the so-called {\sl Mueller-Navelet jets} 
\cite{MN}. The main contribution at the parton level comes 
from gluon exchange in the crossed channel. Then the BFKL theory dresses 
the exchanged gluon with
the multiple emission of gluons, that uniformly fill the rapidity interval 
between the two extreme jets, and resums the leading powers in $\eta$,
including both real and virtual corrections. The signature of the BFKL
dynamics in this context is an exponential rise in $\eta$ of the 
inclusive two-jet cross section. However, since $\hs=x_1x_2s$, with $x_1$
and $x_2$ the momentum fractions of the incoming partons, the
rapidity interval may be increased by keeping $x_1$ and $x_2$ fixed and by 
raising $s$, which may be fulfilled only at a variable-energy collider, or 
by increasing $x_1$ and $x_2$ at a fixed-energy collider. The second option,
though, introduces a damping in the cross section, due to the falling
parton luminosity as $x\ra 1$, which conceals the dynamic rise
of the partonic cross section, induced by the BFKL ladder \cite{DS}. 

\begin{figure}[htb]
\vspace{12pt}
\vskip 0cm
\epsfysize=6cm
\centerline{\epsffile{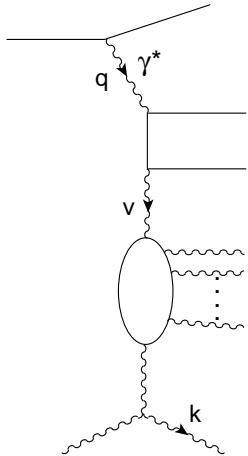}}
\vspace{18pt}
\caption{Forward-jet production in DIS.}
\label{fig:one}
\vspace{12pt}
\end{figure}

In the case of $ep$ colliders \cite{muel}, \cite{bdl}, \cite{kms}
the evolution parameter $\eta$ of the BFKL ladder is 
$\eta =\ln(\hs_/Q^2) = \ln(x/x_{bj})$, 
with $x$ the momentum fraction of the parton initiating the
hard scattering. Since a fixed-energy $ep$ collider is a variable-energy 
collider in the photon-proton frame, it is possible to increase 
$\eta$ by decreasing $x_{bj}$ while keeping $x$ fixed,
thus avoiding the kinematical limitations noticed in the
hadron colliders. 
The lowest-order process featuring gluon exchange in the 
crossed channel is three-parton production at $O(\a^2\a_s^2)$. The exchanged
gluon is then dressed with a BFKL ladder (Fig.\ref{fig:one}). 
In sect.~\ref{sec:one} we reproduce the cross section
for inclusive forward-jet production. As
detailed in the next paragraph, we work out the cross section in the lab 
frame, however our result coincides with the one in the photon-proton frame 
\cite{bdl} \footnote{The relevant
theoretical formulae may also be obtained by taking the massless limit of
heavy-flavor production in DIS \cite{cch}},
since the forward-jet production rate is invariant under boosts 
between the two frames in the high-energy limit.

Another signature of the BFKL dynamics
in Mueller-Navelet jet production in hadron-hadron collisions is
the correlation in the azimuthal angle $\phi$ between the two tagging jets. 
The correlation has a maximum at $\phi=\pi$ and is expected to decrease as 
$\eta$ grows, because of the multiple gluon emission between the 
tagging jets induced by the BFKL ladder \cite{DS,d0}.
The analogous process in DIS is the correlation in $\phi$ between the electron
and the forward jet. Transverse-momentum and azimuthal-angle distributions in 
DIS have been considered previously in the parton model \cite{zee}, \cite{rav} 
and in perturbative QCD at $O(\a_s)$ \cite{geor}, \cite{meng}. In 
the parton model process $e+q \to e+q$, i.e. at $x=x_{bj}$, the 
simple two-body kinematics constrain the jet and
the electron to be produced back-to-back, thus at the parton level it is
$\phi=\pi$, with a smearing induced by the hadronic corrections \cite{zee}, 
\cite{rav}. For the $O(\a_s)$ corrections to the parton-model jet production,
$e+g \to e+q\bar{q}$ or $e+q \to e+q+g$,
the distribution in $\Phi$ in the photon-proton frame has been considered,
with $\Phi$ the azimuthal angle between the jet and the incoming lepton.
It has the functional form,
\beq
{d\s\over d\Phi} \sim A + B \cos{\Phi} + C \cos(2\Phi)\, .\label{how}
\eeq
The $\Phi$-dependence arises from interference of the
photon helicity states $0,+,-$ (longitudinal and transverse polarization).
The term $B \cos{\Phi}$ results from the mixing of 
longitudinal and transverse polarization and
$C\cos(2\Phi)$ occurs when the helicities $\pm$ interfere
\cite{geor}.  In  the simple photon-gluon diagram $C$ 
turns out to be positive \cite{geor}, \cite{meng}.

In this paper, we consider the $O(\a_s^2)$ corrections to forward-jet 
production, $e+g \to e+q\bar{q}+g$ (Fig.1) or $e+q \to q\bar{q}+q$
as a function of the evolution parameter, $\eta =
\ln(x/x_{bj})$, of the BFKL ladder, and compute the distribution in $\phi$
in the lab frame, with $\phi$ the azimuthal angle between the  
electron and the forward jet. We find a correlation in $\phi$ of the form
\beq
{d\s\over d\phi} \sim A' + C' \cos(2\phi)\, ,\label{howy}
\eeq
with $C'$ being negative, now. Compared to the parton-model analysis 
of DIS with $O(\a_s)$ corrections one finds a maximum at $\phi=\pi/2$ 
rather than a minimum. The effect of 
the BFKL ladder is then as usual to flatten the $\phi$ distribution.
The $\cos(\phi)$-term completely drops out due to antisymmetry
in the polar angle distribution ($\theta \rightarrow \pi -\theta$)
of the quark-antiquark pair at the top of the diagram. At $O(\a_s^2)$ 
these quarks are predominantly produced backwards, i.e they are
not tagged as forward jet. After integration over their total phase
space the antisymmetric contributions cancel out.

\section{Forward-jet production}
\label{sec:one}

We work in the electron-proton lab frame and consider the
lowest-order process featuring gluon exchange in the 
crossed channel. In order to achieve that, we must couple the virtual
photon to the off-shell gluon via a quark box. Since we want to examine
jet production near the proton fragmentation region, we couple then the 
off-shell gluon to a parton coming from the proton. The lowest-order
diagrams with these features are the three-parton production one at 
$O(\a^2\a_s^2)$, pictured in fig.\ref{fig:one}, and the one obtained from
this by crossing the quark legs. All the other diagrams at the same order
in $\a_s$ are subleading in the high-energy limit.

We label $p_e$ and $p_{e'}$ the momenta of the incoming and outgoing electrons,
$p$ and $k$ the momenta of the incoming parton and the parton in the forward 
direction, $p_q$ and $p_{\bar q}$ the momenta of the quarks produced in 
the photon-gluon fusion, respectively.
The jet-production rate may be then written in the high-energy limit as
\beq
d\s = \int dx\, f_{eff}(x,\mu) d\hsig\, ,\label{zero}
\eeq
with the effective parton density, $f_{eff}(x,\mu)$, given by \cite{CM}
\beq
f_{eff}(x,\mu) = G(x,\mu) + {4\over 9}\sum_f
\left[Q_f(x,\mu) + \bar Q_f(x,\mu)\right]\, ,\label{effec}
\end{equation}
with the sum over the quark flavors of the incoming parton. $\mu$
denotes the factorization scale.
The partonic cross section, $d\hsig$, is
\beq
d\hsig = {(2\pi)^4 \delta^4(p_{e'}+p_q+p_{\bar q}+k-p_e-p)\over 2xs}\,
d\Pi_k d\Pi_{p_{e'}} d\Pi_{p_q} d\Pi_{p_{\bar q}} |{\cal M}|^2\, ,\label{half}
\eeq
with $s$ the squared electron-proton center-of-mass energy, and with the 
phase space given in terms of rapidity $\eta$ and transverse momentum $\kt$ by,
\beq
d\Pi = {d\eta d^2\kt\over 16\pi^3}\, .\label{ps}
\eeq
We introduce light-cone, or Sudakov, variables, with light-cone directions 
taken to be the ones of the proton and the electron, which we call $+$ and $-$
respectively. In the high-energy limit, for which the momenta $p_{e'}^+$, 
$p_q^+$ and $p_{\bar q}^+$ are negligible with respect to the one of the 
forward jet, $k^+$, the momentum conservation, $\delta(\sum p^+)$, fixes $x$
in eq.(\ref{zero}),
\beq
x={\kt e^{\eta_k}\over 2E_P}\, ,\label{cons}
\eeq
with $E_P$ the proton energy and $\eta_k$ the jet rapidity. With obvious
modifications of the DIS standard conventions we may write
the square of the invariant amplitude, integrated over the phase space
of the outgoing quarks produced in the photon-gluon fusion, as 
\beq
\int d\Pi_{p_q} d\Pi_{p_{\bar q}} |{\cal M}|^2 \delta(\sum p^-) 
\delta^2(\sum p_{\perp}) = {e^4\over Q^4} L_{\mu\nu} 
W_{\mu\nu} \sum_q e_q^2\, ,\label{one}
\eeq
with $L_{\mu\nu}$ and $W_{\mu\nu}$ the tensors which describe the leptonic 
and hadronic structure respectively, and include the average
(sum) over the initial-state (final-state) spin and color degrees of freedom.
On the right-hand side we have singled the sum over the flavors of the 
final-state quarks out of $W_{\mu\nu}$. The leptonic tensor has the form,
\beq
L_{\mu\nu} = 2(p_e^{\mu} p_{e'}^{\nu} + p_{e'}^{\mu} p_e^{\nu} - g_{\mu\nu}
p_e\cdot p_{e'})\, ,\label{two}
\eeq
where the contribution of the $Z$-boson exchange has been neglected. 
The hadronic tensor $W_{\mu\nu}$ depends on the momenta of the proton $P$,
the virtual-photon $q$ and the jet $k$. Thus for an unpolarized
cross section it may be expressed in terms of four structure functions
\cite{rav}. We have found convenient, though, just to determine the 
contractions
$W_{\mu\nu}g^{\mu\nu}$ and $W_{\mu\nu}p_e^{\mu} p_{e'}^{\nu}$, with
$W_{\mu\nu}p_e^{\mu} p_{e'}^{\nu}=W_{\mu\nu}p_e^{\mu} p_e^{\nu}$ because of
gauge invariance. For the gluon coupling to the gluon exchanged in the 
crossed channel (fig.\ref{fig:one}), we use the high-energy limit, i.e. we
retain only the helicity-conserving term \cite{BFKL}. This entails that the
component $v^-$ of the gluon exchanged in the crossed channel is neglected
with respect to $p_q^-$ and $p_{\bar q}^-$.
The calculation is long and tedious, and follows the lines of the ones 
performed in ref.~\cite{bdl}, \cite{kms} and \cite{cch}.
We obtain the forward-jet production cross section,
\beq
{d\s\over d\eta_k d\kt^2 d\eta_e dq_{\perp}^2 d\phi} = x f_{eff}(x,\mu)
{d\hsig\over  d\kt^2 d\eta_e dq_{\perp}^2 d\phi}\, ,\label{mas}
\eeq
with $\eta_e$ the electron rapidity, $q_{\perp}=-p_{e\perp}$,
$\phi$ the azimuthal angle between the jet and the electron,
$x$ fixed by eq.(\ref{cons}), and $f_{eff}$ taken from eq.(\ref{effec}). 
Introducing the electron energy loss
$y$, satisfying the relation $q_{\perp}^2=(1-y)Q^2$, eq.(\ref{mas}) reads:
\beq
{d\s \over dx d\kt^2 dy dQ^2 d\phi}= f_{eff}(x,\mu)
{d\hsig\over  d\kt^2 dy dQ^2 d\phi} \label{jac}\;\;.
\eeq
The partonic cross section in
eq.(\ref{jac}) is
\beq
{d\hsig\over d\kt^2 dy dQ^2 d\phi} = {N_c\a^2\a_s^2\over 2\pi^2}
\sum_q e_q^2 {1\over y (Q^2\kt^2)^2}
{\cal F}(\kt^2,Q^2,\phi,y)\, ,\label{nice}
\eeq
with $N_c=3$ the number of colors. The impact 
factor for the final-state quarks, ${\cal F}(\kt^2,Q^2,\phi,y)$, is
\bea
& & {\cal F}(\kt^2,Q^2,\phi,y) = \kt^2 Q^2 \int_0^1 d\a \int_0^1 dz 
\left[\a(1-\a)Q^2 + z(1-z)\kt^2\right]^{-1} \nn\\
&\times& \left\{\left[{1\over 2} - \a(1-\a) - z(1-z) + 2\a(1-\a)z(1-z)\right]
y^2 \right. \label{fac}\\
&+& \left. \left[1 - 2z(1-z) - 2\a(1-\a) + 12\a(1-\a)z(1-z)\right](1-y)
\right. \nn\\ 
&-& \left. 4\a(1-\a)z(1-z)(1-y)\cos(2\phi )\right\}\, .\nn
\eea
The BFKL ladder we wish to insert on the gluon exchanged in the crossed 
channel is \cite{BFKL},
\beq
f(\kt,v_{\perp},\phi,\tilde{\phi},\eta)\, =\, {1\over (2\pi)^2}\, 
{1\over (\kt^2 
v_{\perp}^2)^{1/2}} \sum_{n=-\infty}^{\infty} 
e^{in(\tilde{\phi}-\phi)}\, \int_{-\infty}^{\infty} d\nu\, 
e^{\omega(\nu,n)\eta}\, e^{i\nu\ln(v_{\perp}^2/\kt^2)}\, ,\label{lad}
\eeq
with $v_{\perp}$ the transverse momentum of the off-shell gluon coupling to
the quark box (Fig.\ref{fig:one}), $\tilde{\phi}$ the azimuthal angle 
between $v_{\perp}$ and ${p_e}_\perp$, $\eta=\ln(x/x_{bj})$ and
\beq
\omega(\nu,n)\, =\, -2{\a_s N_c\over\pi}\, {\rm Re}\left[\psi\left({|n|+1
\over 2} +i\nu\right) -\psi(1)\right]\, ,\label{om}
\eeq
with $\psi$ the logarithmic derivative of the $\Gamma$ function. 
The corrections of the BFKL ladder to the partonic cross section (\ref{nice})
are then given by the formula,
\beq
{d\hsig\over d\kt^2 dy dQ^2 d\phi} = {N_c\a^2\a_s^2\over \pi^2}
\sum_q e_q^2 {1\over y (Q^2)^2 \kt^2}
\int {d^2v_{\perp}\over v_{\perp}^2} f(v_{\perp}^2,\kt^2,\phi, 
\tilde{\phi},\eta) 
{\cal F}(v_{\perp}^2,Q^2,\tilde{\phi},y)\, .\label{comp}
\eeq
In the $\a_s\eta\ra 0$ limit, the 
BFKL ladder reduces to the gluon propagator exchanged
in the crossed channel,
\beq
\lim_{\a_s\eta\ra 0} f(\kt,v_{\perp},\tilde{\phi},\eta)\, =\, {1\over 2} 
\delta^2(\kt-v_{\perp})\, ,\label{born}
\eeq
and accordingly eq.(\ref{comp}) reduces to eq.(\ref{nice}). We term 
eq.(\ref{born}) the Born approximation to the BFKL ladder. We substitute
eq.(\ref{fac}) and (\ref{lad}) into eq.(\ref{comp}), make the measure
$d^2v_{\perp}$ explicit as $d^2v_{\perp}=d\tilde{\phi} dv_{\perp}^2/2$, and
perform the integral over $\tilde{\phi}$, which singles out
the components $n=0,2$ of the eigenvalue (\ref{om}) in the BFKL ladder
(\ref{lad}). We can perform the integral over $\a$, $z$ and $v_{\perp}^2$,
by using the formula
\bea
& &\int_0^1 d\a [\a(1-\a)]^{t_{\a}} \int_0^1 dz [z(1-z)]^{t_z}
\int_0^{\infty} dv_{\perp}^2 {(v_{\perp}^2)^{-1/2+i\nu}\over
\a(1-\a)Q^2 + z(1-z)v_{\perp}^2} =\label{int}\\ & & (Q^2)^{-1/2+i\nu}\, 
{\pi\over\cosh(\pi\nu)}\,
B\left({1\over 2}+t_{\a}+i\nu,{1\over 2}+t_{\a}+i\nu\right)\,
B\left({1\over 2}+t_z-i\nu,{1\over 2}+t_z-i\nu\right)\, ,\nn
\eea
with $t_{\a}=0,1$ and $t_z=0,1$, and with
$B$ the Euler beta function. Performing then a bit of algebra, 
the partonic cross section eq.(\ref{comp}) becomes,
\bea
& & {d\hsig\over d\kt^2 dy dQ^2 d\phi} = \nn\\ & & 
{N_c \a^2 \a_s^2\over 8\pi} \sum_q e_q^2 {1\over (Q^2\kt^2)^{3/2} y}
\int_0^{\infty} d\nu\, \cos\left(\nu\ln{Q^2\over\kt^2}\right)\, 
{\sinh(\pi\nu)\over\cosh^2(\pi\nu)}\, {1\over\nu(1+\nu^2)} \label{jet}\\ 
&\times& \left(e^{\omega(\nu,0)\eta}\,\left[\left(3\nu^2+{11\over 4}\right)
(1-y) + \left(\nu^2+{9\over 4}\right){y^2\over 2}\right]
- e^{\omega(\nu,2)\eta}\, \cos(2\phi)\, \left(\nu^2+{1\over 4}\right)
(1-y)\right)\, .\nn
\eea
%
Substituting eq.(\ref{jet}) into eq.(\ref{jac}), we obtain the forward-jet 
production rate, with the higher-order corrections of the BFKL ladder.

Integrating over $\phi$, the forward-jet production rate reduces
to the one given in ref.\cite{bdl}. In that context it was derived in
the photon-proton frame, however in the high-energy limit the forward-jet 
production rate is invariant under boosts between the electron-proton and 
the photon-proton frames, i.e. in the notation of eq.(\ref{how}) $\Phi=\phi$.

Assuming $\kt^2$ to be of the order $Q^2$ and $\eta$ to be very large 
the integration in eq.(\ref{jet}) can be performed approximately by
means of the saddle point method. After expanding $\omega(\nu,0)$ around 
$\nu=0$ one finds
\beq
{d\hsig\over d\kt^2} \sim {1\over (\kt^2)^{3/2}}\, 
\exp \left( 4 \ln2 \frac{N_c \alpha_s}{\pi} \eta \right)
 \,\exp\left(-{\ln^2(\kt^2/
Q^2)\over 4B\eta}\right) \qquad {\rm with} \quad B=14\zeta(3){N_c\a_s\over\pi}
\, ,\label{asymp}
\eeq
where $\zeta$ denotes the Riemann-$\zeta$ function. Eq.(\ref{asymp}) is 
regular as 
$\kt\ra 0$, however the approximation becomes invalid then.
The limit $\kt\ra 0$ is of interest for the total,
integrated cross section which is dominated by small $\kt$.
Assuming $\log(Q^2/\kt^2)\ra \infty$ the saddle point of 
eq.(\ref{jet}) is shifted towards the pole of $\omega(\nu,0)$ at
$\nu=-i/2$. This pole corresponds to the collinear emission of gluons,
i.e. we pick up all collinear singularities of the gluon ladder.
The gluon structure function according to its definition includes all collinear
singularities, so that after integration over the jet momentum 
eq.(\ref{jet}) coincides at least in a formal sense
with the inclusive small $x_{bj}$ cross section.

Returning to the distribution in $\phi$ of eq.(\ref{jet}) we recognize
that it is periodic in $\pi$
and has a maximum at $\phi=\pi/2$. The electron-jet correlation in $\phi$,
though, quickly dies out as $\eta=\ln(x/x_{bj})$ increases. Indeed, the
main contribution to the integral over $\nu$ in eq.(\ref{jet}) comes from
the $\nu\simeq 0$ region, and from eq.(\ref{om}) we see that
$\omega(\nu=0,n=0)= 4\ln{2}(\a_s N_c/\pi)$ while
$\omega(\nu=0,n=2)= 4(\ln{2}-1)(\a_s N_c/\pi)$, thus the uncorrelated
term in eq.(\ref{jet}) is enhanced while the correlated one is upset as 
$\eta$ increases. 

For a small $\eta=\ln(x/x_{bj})$ the zeroth order
parton configuration is relevant, and the usual correlation at $\phi=\pi$,
between the electron and the current-jet
is obtained. As $\eta$ grows the jet production is increasingly
dominated by diagrams with two- and later on with three-final state partons 
and with gluon exchange in the crossed channel. The three parton final
state has the functional form of eqs.(\ref{howy}) or
(\ref{nice}). The higher-order corrections to them induced by the BFKL 
ladder, eq.(\ref{jet}), dampen then the correlation in $\phi$.
What said above does not suffice, though, to explain the correlation
at $\phi=\pi/2$. In order to see that and the
transition from the correlation at $\phi=\pi$ to the one at
$\phi=\pi/2$, it is necessary to compare the asymptotic calculation of
eq.(\ref{nice}) with the exact calculation 
at $O(\a_s^2)$ \cite{us}.

Finally, we note that because of the relation between the DIS
cross section and the structure functions $F_{1(2)}$,
\beq
{d\s\over dy dQ^2} = {4\pi \a^2\over y Q^4}\, \left[(1-y) F_2(x_{bj},
Q^2) + x_{bj} y^2 F_1(x_{bj}, Q^2)\right]\, ,\label{otto}
\eeq
with $F_1$ and $F_2$ related to the transverse and longitudinal
polarizations of the virtual-photon total cross section by,
\bea
F_1(x_{bj}, Q^2) &=& {Q^2\over 8\pi^2\a x_{bj}} \s_T(\gamma^* P)\, 
,\label{cro}\\
F_2(x_{bj}, Q^2) &=& {Q^2\over 4\pi^2\a} \left(\s_T(\gamma^* P) + 
\s_L(\gamma^* P)\right)\, ,\nn
\eea
in the high-energy limit all the information about the correlation in $\phi$ 
is effectively contained in the expression for the 
longitudinal polarization of the virtual 
photon although it originates from the interference of the transverse
helicities. Indeed,
comparing the differential of eq.(\ref{otto}) in $\phi$ to eq.(\ref{jac}),
with the partonic cross section given by eq.(\ref{nice}) or (\ref{jet}), we
see that $dF_2/d\phi$ depends on $\phi$, while $dF_1/d\phi$ does not.

\section*{Acknowledgements}
 
VDD wishes to thank Stefano Catani for useful discussions.

\end{document}